\documentclass[hyper,a4paper]{JHEP3}
\usepackage{multirow}

\usepackage[dvips]{graphicx}
\usepackage{epsfig}
\usepackage{rotating}
\usepackage{amssymb}
\usepackage{amsmath}

\newcommand{\lsim}{\mathrel{\mathop{\kern 0pt \rlap
      {\raise.2ex\hbox{$<$}}}\lower.9ex\hbox{\kern-.190em $ \sim$}}}

\newcommand{\gsim}{\mathrel{\mathop{\kern 0pt
      \rlap{\raise.2ex\hbox{$>$}}}\lower.9ex\hbox{\kern-.190em $\sim$}}}

\newcommand{\beq}{\begin{equation}}
\newcommand{\eeq}{\end{equation}}

\newcommand{\be}{\begin{equation}}
\newcommand{\ee}{\end{equation}}
\newcommand{\beqarr}{\begin{eqnarray}}
\newcommand{\eeqarr}{\end{eqnarray}}

\preprint{KIAS-P12036}

\title{Non-perturbative Effect and PAMELA Limit on Electro-Weak Dark Matter}

\author{Eung Jin Chun  and Jong-Chul Park \\
\\Korea Institute for Advanced Study\\
Heogiro 87, Dongdaemun-gu\\
Seoul 130-722, Korea\\
Emails: \email{ejchun@kias.re.kr, jcpark@kias.re.kr} }
\author{Stefano Scopel \\
Department of Physics, Sogang University\\
1 Sinsu-dong, Mapo-gu\\
        Seoul 121-742, Korea\\
Email: \email{scopel@sogang.ac.kr} }

\abstract{ We discuss the non-perturbative effects on the annihilation
  cross section of an Electro-Weak Dark Matter (EWDM) particle
  belonging to an electroweak multiplet when the splittings between
  the masses of the DM component and the other charged or neutral
  component(s) of the multiplet are treated as free parameters.  Our
  analysis shows that EWDM exhibits not only the usual Sommerfeld
  enhancement with resonance peaks but also dips where the cross
  section is suppressed.  Moreover, we have shown that the
  non--perturbative effects become important even when the EWDM mass
  is below the TeV scale, provided that some of the mass splitting are
  reduced to the order of a few MeV.  This extends the possibility of
  observing sizeable non-perturbative effects in the dark matter
  annihilation to values of the dark matter mass significantly smaller
  than previously considered, since only electroweak--induced mass
  splittings larger than 100 MeV have been discussed in the literature
  so far.  We have then used the available experimental data on the
  cosmic antiproton flux to constrain the EWDM parameter space. In our
  calculation of the expected signal we have included the effect of
  the convolution of the cross section with the velocity distribution
  of the dark matter particles in the Galaxy, showing that it can
  alter the non--perturbative effects significantly.  In the case of
  EWDM with non-zero hypercharge, we have shown that the mass
  splitting in the Dirac dark matter fermion can be chosen so that the
  inelastic cross section of the EWDM off nuclei is allowed by present
  direct detection constraints and at the same time is within the
  reach of future experiments.}  \keywords{Dark Matter, Sommerfeld effect, Ramsauer-Townsend effect}

\maketitle

\begin{document}


\section{Introduction}
A multiplet of the Standard Model gauge group $SU(2)_L\times
U(1)_Y$ \cite{hisano03} would be the simplest candidate of dark
matter as it requires no additional `ad hoc' interaction or
couplings. Its electroweak interaction leads to a right thermal
relic density if its mass is in the multi TeV range, and its
stability is guaranteed automatically for a certain higher
dimensional multiplet \cite{minimal_dm}.  Another important
feature of such an ``Electro-Weak Dark Matter'' (EWDM) is
non-perturbative effects on its annihilation into gauge bosons
which modifies significantly the tree-level results in the
determination of its thermal relic density and its indirect
detection rate \cite{hisano03,cirelli07}.

In this paper, we study such a non-perturbative correction for a
Majorana or Dirac fermion EWDM in a wide range of its mass. That is,
we will consider a ``non-minimal'' EWDM allowing an unspecified
non-standard cosmology for the generation of a right relic density and
a certain discrete symmetry for its stability.  In the present
Universe, the dark matter is highly non-relativistic and thus the
wave-functions of the effective non-relativistic two-body EWDM states
can get strongly modified by the non-abelian electroweak(EW) potential
in the process of their pair annihilation \cite{hisano03}. This is a
generalization of the ``Sommerfeld effect'' \cite{sommerfeld}
well-known for a single-component dark matter carrying an abelian
gauge charge.  Recall that the non-perturbative correction enhances
(reduces) the annihilation cross section in case of an attractive
(repulsive) Coulomb potential \cite{cirelli07}. It will be interesting
to note that an EWDM exhibits not only the usual Sommerfeld
enhancement and resonance peaks but also a suppressed cross section
for a certain choice of the parameters. This is a realization of the
``Ramsauer-Townsend (RT) effect'' \cite{ramsauer} observed in
low-energy electron scattering off gas atoms.  We will show that such
effects are caused mainly by the electromagnetic(EM) interaction in
the two-body states of the charged components of an electroweak
multiplet. An important feature is that the velocity distribution of
dark matter particles has to be included in the non-perturbative
calculation of the EWDM annihilation rate, as the
``Sommerfeld-Ramsauer-Townsend'' (SRT) resonance effect occurs when a
certain condition is met among the model parameters, including the
kinetic energy and so the speed of the annihilating states.

The non-perturbative correction to the EWDM annihilation cross section
is important to set a limit on the EWDM, the most stringent
constraints coming from the PAMELA measurement of the cosmic
anti-proton flux \cite{pamela2008,pamela2010} and from the recent
FERMI-LAT measurement of the diffuse gamma-ray emission in dwarf
spheroidal galaxies \cite{fermi11}.  In the present paper we will
focus on the PAMELA antiproton limit, deriving a constraint which is a
slightly stronger than the FERMI--LAT limit, and comparable to the
result of Ref.~\cite{belanger12} with the `MED' propagation
astrophysical parameters and a fixed secondary background.

In Section \ref{General-EWDM}, we will give a general description of
various electroweak multiplet dark matter candidates and present
formulae to calculate the non-perturbative effect, summarizing the
results in Refs.~\cite{hisano03,cirelli07}.  Depending on whether the
EWDM carries a hypercharge or not, it can be a Dirac or a Majorana
fermion. In the first case, the neutral Dirac components have to split
into two Majorana fermions with a mass gap sufficient to suppress the
(inelastic) nucleonic scattering through the exchange of a $Z$
boson. In Section \ref{sec:srt}, we discuss various features of the
non-perturbative correction showing the SRT effect for the simplest
example of a triplet EWDM with no hypercharge. Section
\ref{sec:direct} discusses values of the mass splitting which are
large enough to avoid the current direct detection bound but still
detectable in future experiments, depending on the dark matter
mass. The antiproton yield from the EWDM annihilation to gauge bosons
is analyzed in Section \ref{sec:pbar} to place a limit from the
current cosmic antiproton flux measurements. We then calculate
non-perturbative annihilation cross sections for various EWDM
candidates to put a mass limit from the PAMELA data on the antiproton
flux in Section 6. We conclude in Section \ref{sec:conclusions}.

\section{General EWDM and non-perturbative effect}
\label{General-EWDM}

The dark matter particle can be the neutral component of an $SU(2)_L
\times U(1)_Y$ fermion multiplet.  As a specific example, we will
consider a vector-like (Dirac) doublet with $Y=\pm 1/2$
(Higgsino-like), a (Majorana) triplet with $Y=0$ (wino-like) and a
vector-like (Dirac) triplet with $Y=\pm1$. Note that a certain
symmetry like $Z_2$ has to be imposed for the stability of these EWDM
candidates.  Furthermore, the dark matter component of a Dirac
multiplet is charged under $U(1)_Y$ and thus its scattering
cross section with nuclei through $Z$ boson exchange is far above the
current limit from direct detection experiments. This constraint is
however invalidated if there is a mass splitting in the Dirac dark
matter fermion and thus the heavier Majorana fermion component cannot
be excited by the nucleonic scattering of the lighter one (assumed to
be the dark matter).  A detailed analysis will be presented later to show
that a mass splitting of order 0.2 MeV would be detectable while still allowed by
the current data.  Such a mass splitting can come from a higher
dimensional operator between the EWDM and the Higgs doublet. For
instance, the Higgsino-like dark matter multiplet, denoted by $\chi_u
= (\chi^+, \chi^0)$ and $\chi_d= (\chi^0_d, \chi^-_d)$ in the chiral
representation, allows the dimension-four operators:
\begin{equation}
 {1\over \Lambda} (H_u \chi_d)^2, \quad {1\over \Lambda} (H_d \chi_u)^2,\quad
 {1\over \Lambda} (H_u \chi_d) (H_d \chi_u),
\end{equation}
where $H_{u}=(H^+, H^0)$ and $H_d=\epsilon H_u^*$ represent the
Higgs doublets coupling to the up and down type quarks,
respectively. Similarly, for the triplet EWDM multiplet with
$Y=\pm1$ consisting of two chiral fermions, $\chi_u=(\chi_u^{++},
\chi_u^+, \chi_u^0)$ and $\chi_d=(\chi_d^{0}, \chi_d^{-},
\chi_d^{--})$, the mass splitting between the Dirac pair
$\chi_{u,d}^0$ can arise from:
\begin{equation}
 {1\over \Lambda^3} (H_u H_u \chi_d)^2, \quad {1\over \Lambda^3} (H_d  H_d \chi_u)^2,\quad
 {1\over \Lambda^3} (H_u H_u \chi_d) (H_d H_d \chi_u) .
\end{equation}
On the other hand, the wino-like EWDM multiplet, a triplet with
$Y=0$ denoted by $\chi=(\chi^+, \chi^0, \chi^-)$ has only one
Majorana neutral component.  Note that a mass splitting between
the charged and neutral components of order 0.1 GeV arises from
the electroweak one-loop correction \cite{minimal_dm}.  In the
following sections, we will assume arbitrarily small mass gaps
among the multiplet components which make a big impact on the
non-perturbative annihilation rate of the EWDM particle.

\medskip

In the non-relativistic pair annihilation of the EWDM, the
non-perturbative effect due to the exchange of the electroweak
gauge bosons mixes together the two-body states of the multiplet
components. In the case of the Higgsino-like EWDM, there are three
states formed by the charged (Dirac) component and two neutral
(Majorana) components: $\chi_u^+ \chi_d^-$, $\chi_1^0 \chi_1^0$
and $\chi_0^0 \chi_0^0$, where $\chi_0^0$ denotes the dark matter
component.  For the wino-like EWDM, we have two two-body states:
$\chi^+ \chi^-$ and $\chi^0\chi^0$. The triplet EWDM with $Y=\pm1$
has four two-body states: $\chi_u^{++}\chi_d^{--}$, $\chi_u^+
\chi_d^-$, $\chi_1^0 \chi_1^0$ and $\chi_0^0 \chi_0^0$.

The Green's functions $g_{ij}$ corresponding to the processes
summarized above, where the indices $i$ and $j$ run over the
two-body states of each EWDM candidate, verify the Schr\"odinger
equation \cite{hisano03}:
\begin{equation} \label{schroedinger}
 -{1\over m_{DM}} {\partial^2 g_{ij} (r) \over \partial r^2} + V_{ik}(r) g_{kj} (r) = K g_{ij}(r),
\end{equation}
with $m_{DM}$ the mass of the dark matter particle, and the boundary
condition $g_{ij}(0)=\delta_{ij}$ and $\partial g_{ij}(\infty)
/\partial r = i \sqrt{m_{DM} (K-V_{ii}(\infty))}g_{ij}(\infty)$. Here
$K=m_{DM} \beta^2$ is the total kinetic energy of the two initial dark
matter particles in the annihilation process, where $\beta$ is the
velocity of the DM particle in the frame of the galactic halo. Then,
the dark matter pair annihilation cross section is given by:
\begin{equation} \label{sigmaDM}
 \sigma v ( \chi^0_0 \chi^0_0 \to A B) = 2 d_{0i} d^*_{0j} \Gamma_{ij}^{AB},
\end{equation}
where $d_{0j} = g_{0j}(\infty)$ and $v=2 \beta$ is the relative
velocity between the two incident DM particles. Here $A,B$ run
over the gauge bosons $(W^+, W^-, Z, \gamma)$, that is, the gauge
boson final states are $AB=(W^+ W^-, ZZ, \gamma Z, \gamma\gamma)$.
Taking the normalization of the covariant derivative $D_\mu =
\partial_\mu + i g_2 A_\mu T^A$ for each gauge boson $A$, the
potential matrix in Eq.~(\ref{schroedinger})  and the tree-level
annihilation matrix $\Gamma_{ij}$ are given by \cite{cirelli07}:
\begin{eqnarray}
&& V_{ij}(r) = 2 \,\delta m_{i0}\, \delta_{ij} - \alpha_2 N_i N_j
 \sum_{A} \left[ T^A_{ij}\right]^2 {e^{-m_A r} \over r},
\end{eqnarray}
where $\delta m_{i0} = m_{\chi_i} - m_{\chi_0}$, and:
\begin{eqnarray}
&&  \Gamma_{ij}^{AB} = {\pi \alpha_2^2 \over 2(1+\delta_{AB}) m^2_{DM}} f(x_A,x_B)
 N_i N_j \left\{ T^A, T^B \right\}_{ii}
  \left\{ T^A, T^B \right\}_{jj}\,, \\
&& \mbox{where}\;\; f(x_A,x_B) \equiv { \left(1-{x_A+x_B \over 2}\right) \over
 \left( 1 - {x_A+x_B \over 4} \right)^2 }
 \sqrt{ 1 - {x_A+x_B \over 2} +{( x_A - x_B)^2 \over 16}} \; \mbox{with}\;
 x_A = {m_A^2 \over m^2_{DM} } \,.\nonumber
\end{eqnarray}
Here the normalization factor $N_i$ is  $1$ or $ \sqrt{2}$ for the
Dirac (charged) or Majorana (neutral) two-body state,
respectively.

\section{Sommerfeld-Ramsauer-Townsend effect}
\label{sec:srt}

In this section, we present a detailed study of the
non-perturbative effect on the EWDM annihilation cross section
$\sigma v^{WW+ZZ} \equiv \sigma v^{WW} + \sigma v^{ZZ}$ including
both final states $W^+W^-$ and $ZZ$. The wino-like EWDM system has
the smallest number of states and parameters: two bound states
($\chi^+\chi^-$ and $\chi^0 \chi^0$) and one mass gap between
them. For this reason, to simplify our discussion in this section
we will focus on the example of wino-like dark matter, unless
otherwise stated. As will be discussed in Section \ref{sec:pbar},
EWDM can copiously produce $W$ and $Z$ bosons leading to a
sizable contribution to the antiproton flux measured by cosmic-ray
detection experiments such as PAMELA. This puts strong constraints
on the masses of the EWDM particles, as will be analyzed in
Section \ref{PAMELA-limit}.

One of the key observations in this work is that the non-perturbative
effect on the EWDM annihilation cross section includes not only the
Sommerfeld effect \cite{sommerfeld} which induces both an overall
enhancement of the cross section and resonance peaks for particular
values of masses and couplings\cite{hisano03}, but also a suppression
or resonance dips. We will refer to this as the ``Ramsauer-Townsend
effect'' in the DM annihilation processes. The Ramsauer-Townsend
effect is a quantum mechanical phenomenon found in the scattering of
electrons by noble gas atoms: the collision probability reaches a
minimum when the electron kinetic energy take a certain value
\cite{ramsauer}. This is analogous to what happens to the transmission
coefficient of a one-dimensional potential well, which is enhanced
(corresponding to a vanishing reflection probability) when certain
conditions are met between the kinetic energy of the incident particle
and the potential depth and width.  A similar phenomenon can occur in
the process of EWDM annihilation in presence of a non-perturbative
electroweak potential.  In order to see how such a Ramsauer-Townsend
resonance arises in this system, we will perform a numerical analysis
of the Schr\"odinger equation (\ref{schroedinger}) by changing the
mass gaps, the electromagnetic, $W$ and $Z$ potentials, and the
velocity of the DM particle.  In particular, the latter is an
important factor which can drastically change the behavior of the
resonance peaks and dips.  For this reason, in Section
\ref{sec:pamela_limits} we will include in our calculation of the
annihilation cross section a convolution over the Galactic velocity
distribution of the impinging dark matter particles, in order to
compare it to the experimental bound.

We mention here that a plot showing dips for particular values of
masses and couplings in the annihilation cross section of Dark
Matter particles interacting in a non--perturbative electroweak
potential was presented in Ref.~\cite{cirelli07}. To our
knowledge, this is the only instance where the Ramsauer-Townsend
effect has been shown in the literature in the context of dark
matter annihilation. However, the authors of Ref.~\cite{cirelli07}
did not mention this effect in their discussion.

\subsection{Dependence on $\delta m_+$}

\begin{figure}
\begin{center}<
\includegraphics[width=0.65\linewidth]{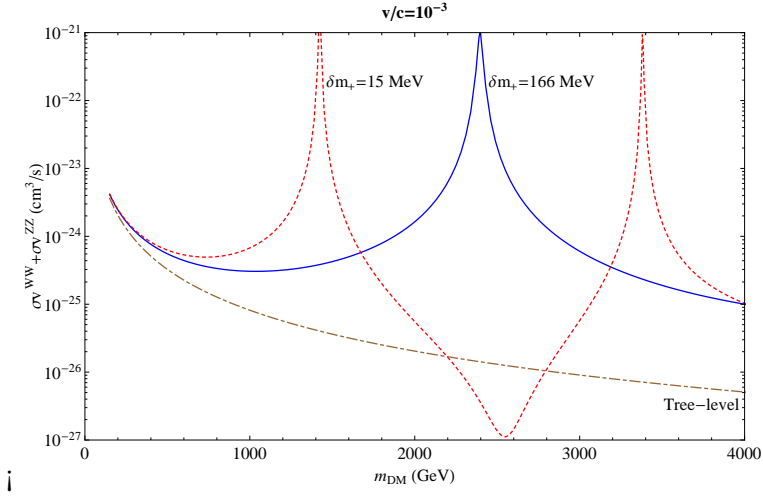}
\end{center}
\caption{Annihilation cross sections  of the wino-like EWDM for
the mass splitting $\delta m_+ \equiv m_{\chi^+}-m_{\chi^0} =$166
MeV (blue solid) and 15 MeV (red dotted). The brown dot-dashed
line shows the tree-level cross section.} \label{Fig_5-1}
\end{figure}

In the determination of the non-perturbative effects of EWDM
annihilation, the splitting between the masses of the dark matter
and charged states is a crucial factor, since it controls the
transition of the DM state to a particle able to experience the
long-range effect of the electromagnetic(EM) interaction.  The
wino-like EWDM has two states $\chi^\pm$ and $\chi^0$ whose mass
splitting is defined by $\delta m_+ \equiv m_{\chi^+}-m_{\chi^0}$.
In Figure \ref{Fig_5-1}, we present the cross sections of the
wino-like DM annihilation to the $W^+W^-$ and $ZZ$ final states
$\sigma v^{WW+ZZ}$, as a function of $m_{\rm DM}$ for the two
representative values $\delta m_+ =$ 166 and 15 MeV, where 166 MeV
is the typical mass splitting arising from the EW one-loop
correction \cite{minimal_dm}. The velocity of the EWDM is fixed to
be the typical value of $v/c=10^{-3}$.  Our result with $\delta
m_+=166$ MeV, which shows the Sommerfeld enhancement and a
resonance peak within the mass range presented, is consistent with
Ref.~\cite{hisano03}. For the smaller mass gap, one finds that the
peak positions shift to smaller $m_{\rm DM}$ and a dip appears
that before was missing. In this latter case, the smaller mass gap
allows an easier access to the charged state, inducing a stronger
non-perturbative effect which activates the Ramsauer-Townsend
suppression.  Actually, in the next subsection we will confirm
that the Ramsauer-Townsend dips appear mainly due to the
electromagnetic interaction of the charged states.

\subsection{Dependence on the EW interactions}

\begin{figure}
\begin{center}
\includegraphics[width=0.65\linewidth]{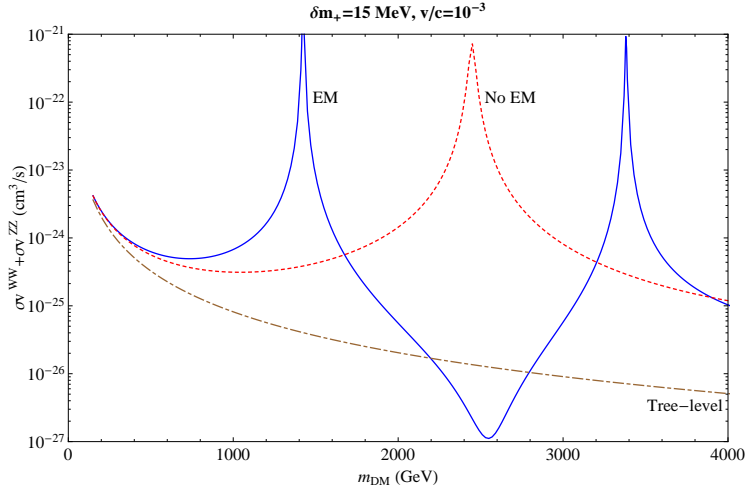}
\end{center}
\caption{Annihilation cross sections  for the wino-like EWDM with
$\delta m_+ =$ 15 MeV. The solid blue and red dotted lines show
the results with and without the EM interaction, respectively. The
brown dot-dashed line is the tree-level cross section.}
\label{Fig_5-2}
\end{figure}

\begin{figure}
\begin{center}
\includegraphics[width=0.49\linewidth]{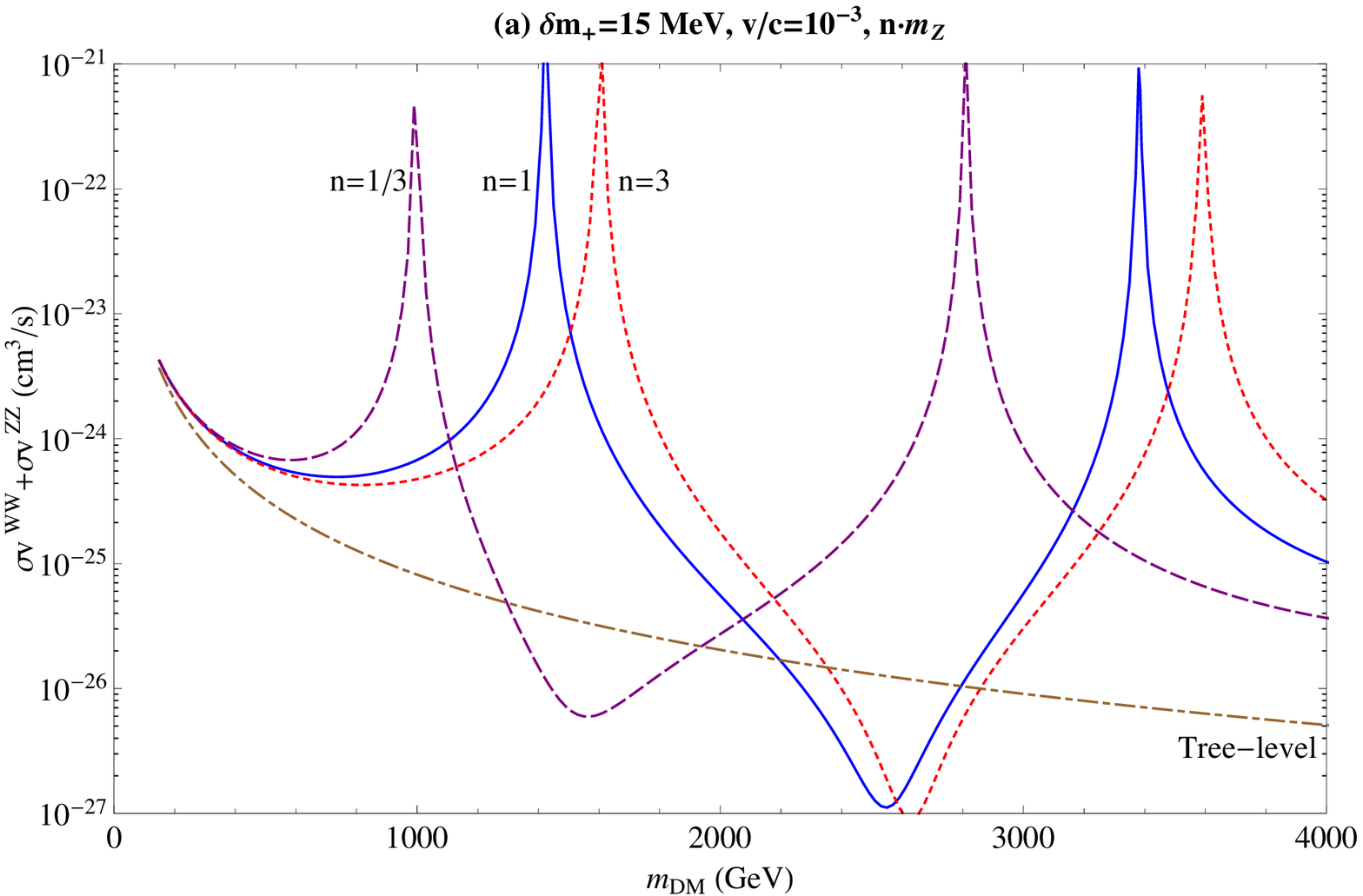}
\includegraphics[width=0.49\linewidth]{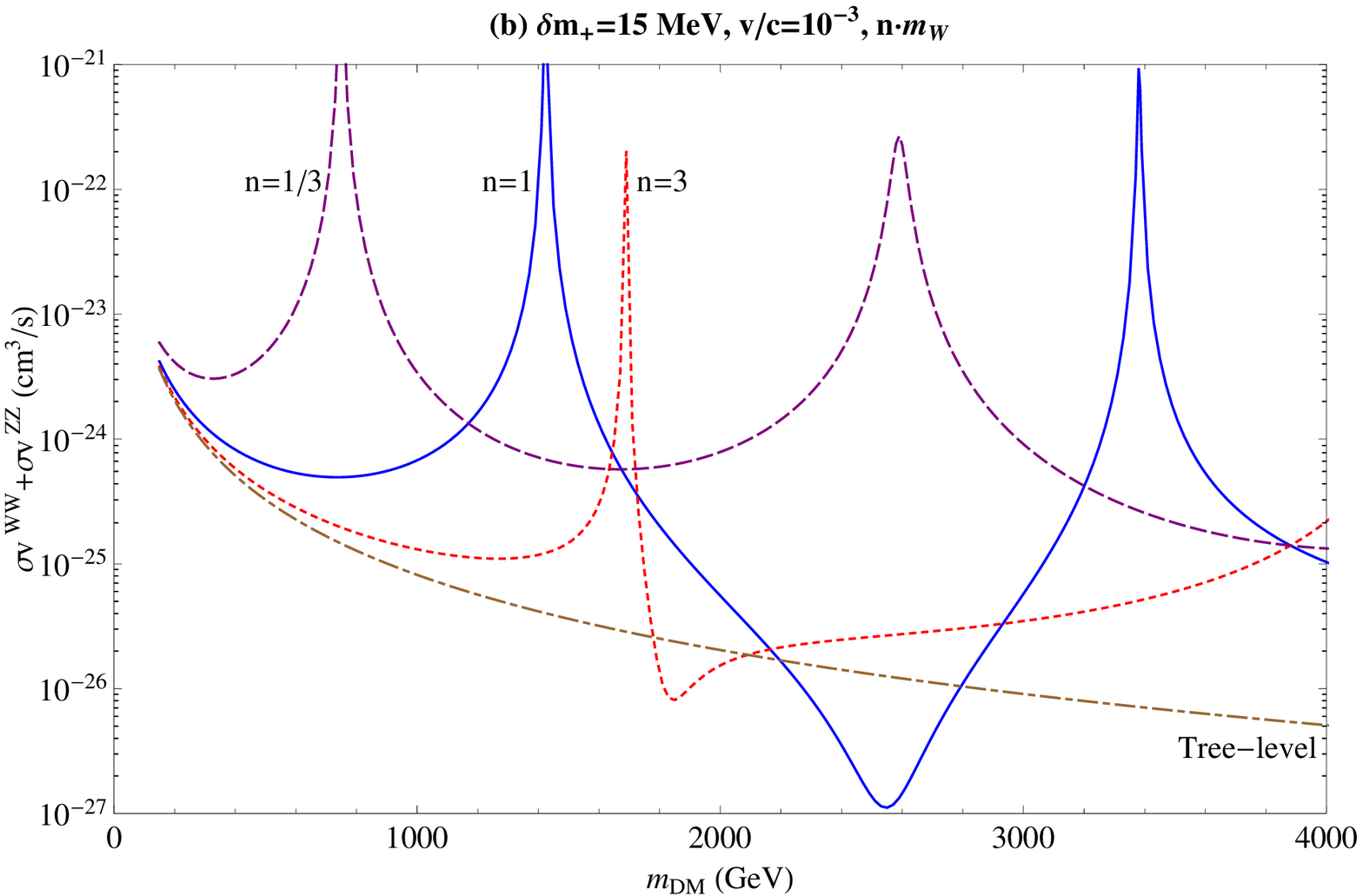}
\end{center}
\caption{Annihilation cross sections  for the wino-like EWDM with
$\delta m_+ =$ 15 MeV. Each panel shows the dependence on the
interactions via the (a) $Z$ and (b) $W$ bosons. The purple
dashed, blue solid, and red dotted lines respectively correspond
to the cases of (a) $n\cdot m_Z$  and (b) $n\cdot m_W$ with
$n=1/3, 1,$ and 3. The brown dot-dashed line is the tree-level
cross section.} \label{Fig_5-3}
\end{figure}

In this subsection, we will modify the electroweak potentials in
order to see how the SRT effect is affected.  In
Figure~\ref{Fig_5-2}, we first plot the annihilation cross
sections to $W^+ W^-$ and $ZZ$ as a function of $m_{\rm DM}$ for
$\delta m_+ =$ 15 MeV with and without the EM interaction. Then in
Figure~\ref{Fig_5-3}, we display the annihilation cross section
for the same parameters of the previous figure, but we vary the
masses of the $Z$ or $W$ boson. In particular, in
Figure~\ref{Fig_5-3}(a) we take $m_Z\rightarrow n\cdot m_Z$ and in
Figure~\ref{Fig_5-3}(b) $m_W\rightarrow n\cdot m_W$, where in both
cases $n=1/3, 1,$ and 3.

Figure \ref{Fig_5-2} shows that the dip disappears when the EM
interaction is turned off, confirming the role played by the EM
potential in the Ramsauer-Townsend effect when the mass gap
between the dark matter and charged states is small. Furthermore,
one finds that the behavior of the annihilation cross section
without the EM interaction is almost the same as in Figure
\ref{Fig_5-1}, where the influence of the EM interaction is
blocked by a larger mass gap. This tells us that the SRT effect is
insensitive to the mass gap as far as there is no long-range EM
interaction.

The weak interaction via the $Z$ or $W$ boson exchange behaves like a
long range interaction for a large DM mass, as can be seen in the
large mass limit of Figures~\ref{Fig_5-1} and \ref{Fig_5-2}, where a
sizeable SRT enhancement is present.  A similar effect is expected to
occur if smaller $Z$ or $W$ boson masses are taken. However, this may
change the resonance conditions for the SRT peaks and dips as well.
As a consequence, Figure \ref{Fig_5-3} shows that the peaks move to
smaller values of the DM mass for smaller $Z$ and $W$ boson masses.
Furthermore, one can find that the SRT effect is more sensitive to a
change of the $W$ boson mass compared to that of the $Z$ boson
mass. This can be explained by the fact that a change of the $W$ boson
mass influences not only the strength of the weak interaction via the
$W$ boson propagator, but also the accessibility thorough an EW
charged current to a state able to interact
electromagnetically. Moreover, Figure \ref{Fig_5-3} (b) shows a
drastic change in the Ramsauer-Townsend effect: in particular, the RT
resonance condition cannot be met for a light $W$ boson mass.

From the above discussion, it is also expected that the SRT peak and
dip positions move to lighter DM masses as the electroweak interaction
strength is increased.  This is indeed what happens in the case of the
quintuplet EWDM discussed in Ref.~\cite{cirelli07}, which shows a RT
dip at around $m_{DM} = 2$ TeV for a mass gap of order 100 MeV. On the
other hand, no RT effect for such large values of the mass gap is
observed for lower--dimensional EWDM and a dark matter mass within the
multi-TeV range.

\subsection{Amplitudes of the wave functions}

\begin{figure}
\begin{center}
\includegraphics[width=0.65\linewidth]{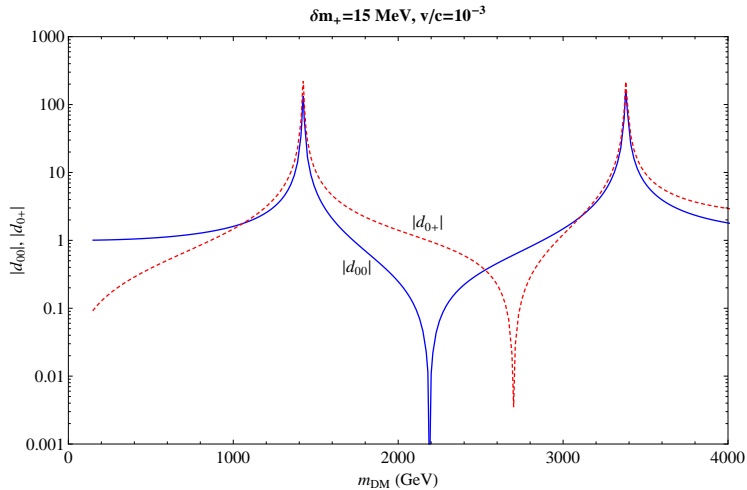}
\end{center}
\caption{Wave function amplitudes for the two-body states of the
wino-like EWDM with $\delta m_+ =$ 15 MeV. The blue solid and red
dotted lines show $|d_{00}|$ and $|d_{0+}|$, respectively.} \label{Fig_5-4}
\end{figure}

As can be seen in Figures \ref{Fig_5-1}, \ref{Fig_5-2}, and
\ref{Fig_5-3}, the annihilation cross sections of the EWDM show peaks
and dips due to the SRT effect.  To see their behavior in more detail,
we present in Figure \ref{Fig_5-4} the amplitudes of the wave
functions $d_{00}$ and $d_{0+}$ connecting the two bound states
$\chi^0\chi^0$ and $\chi^+ \chi^-$ [see Eqs.~(\ref{schroedinger}) and
(\ref{sigmaDM})] of the wino-like EWDM with $\delta m_+ = 15$ MeV.
One finds that each wave function has peaks and dips, implying both a
constructive and a destructive resonance behavior.  However, some
difference is observed in the behavior of the peaks and dips. In
particular, while peak positions coincide in the two wave function
amplitudes, the dip positions do not. As a consequence, the dip in the
annihilation cross section (the red-dotted line in
Figure~\ref{Fig_5-1}) appears between the two dips in the wave
functions (the two lines in Figure~\ref{Fig_5-4}) and is
broader. However, the dip in the cross section can be more pronounced
and even as narrow as in the case of Sommerfeld peaks in situations
where the dips in the two wave functions are closer.

\subsection{Dependence on DM velocity}
\label{velocity}

\begin{figure}
\begin{center}
\includegraphics[width=0.49\linewidth]{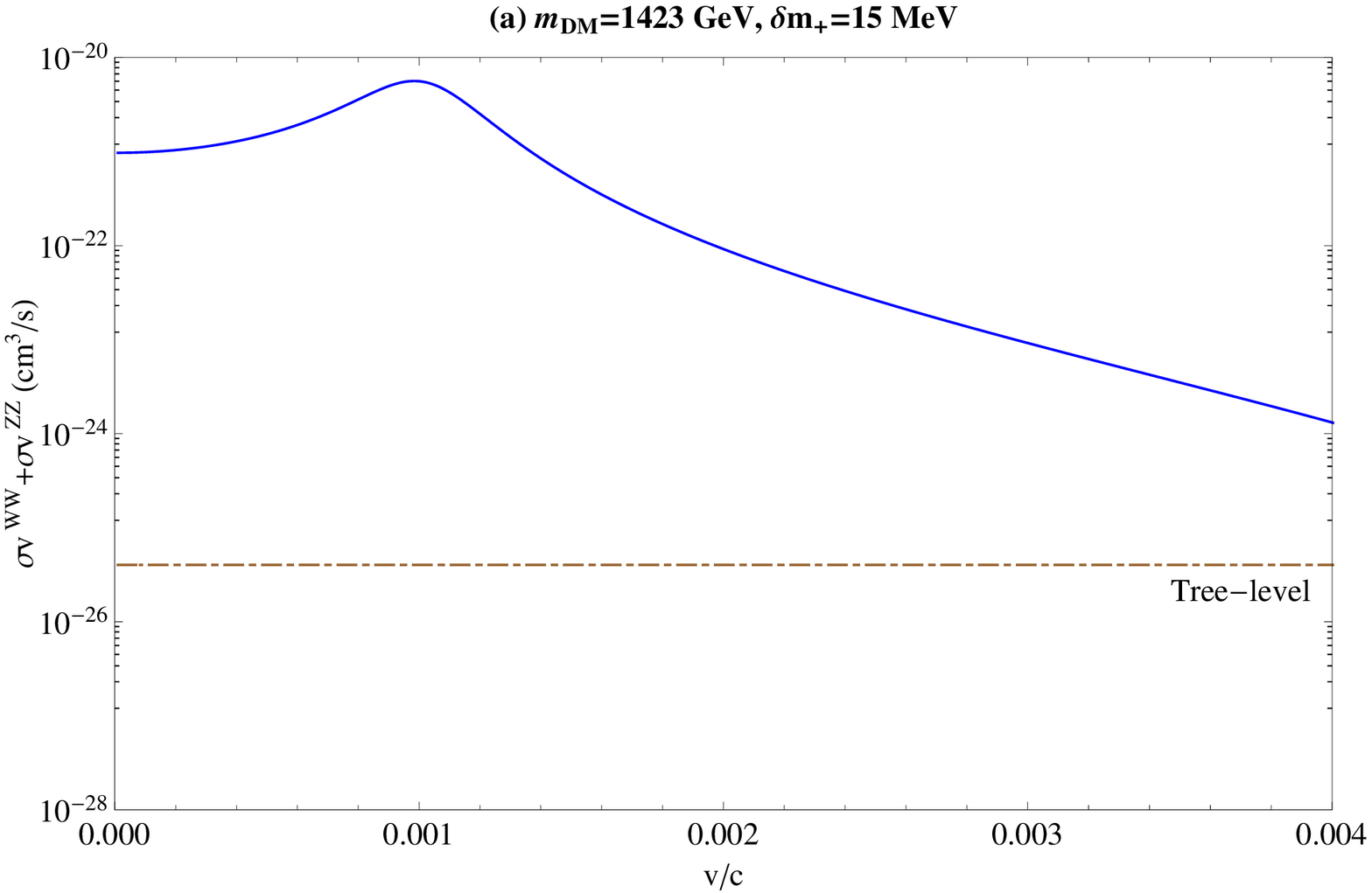}
\includegraphics[width=0.49\linewidth]{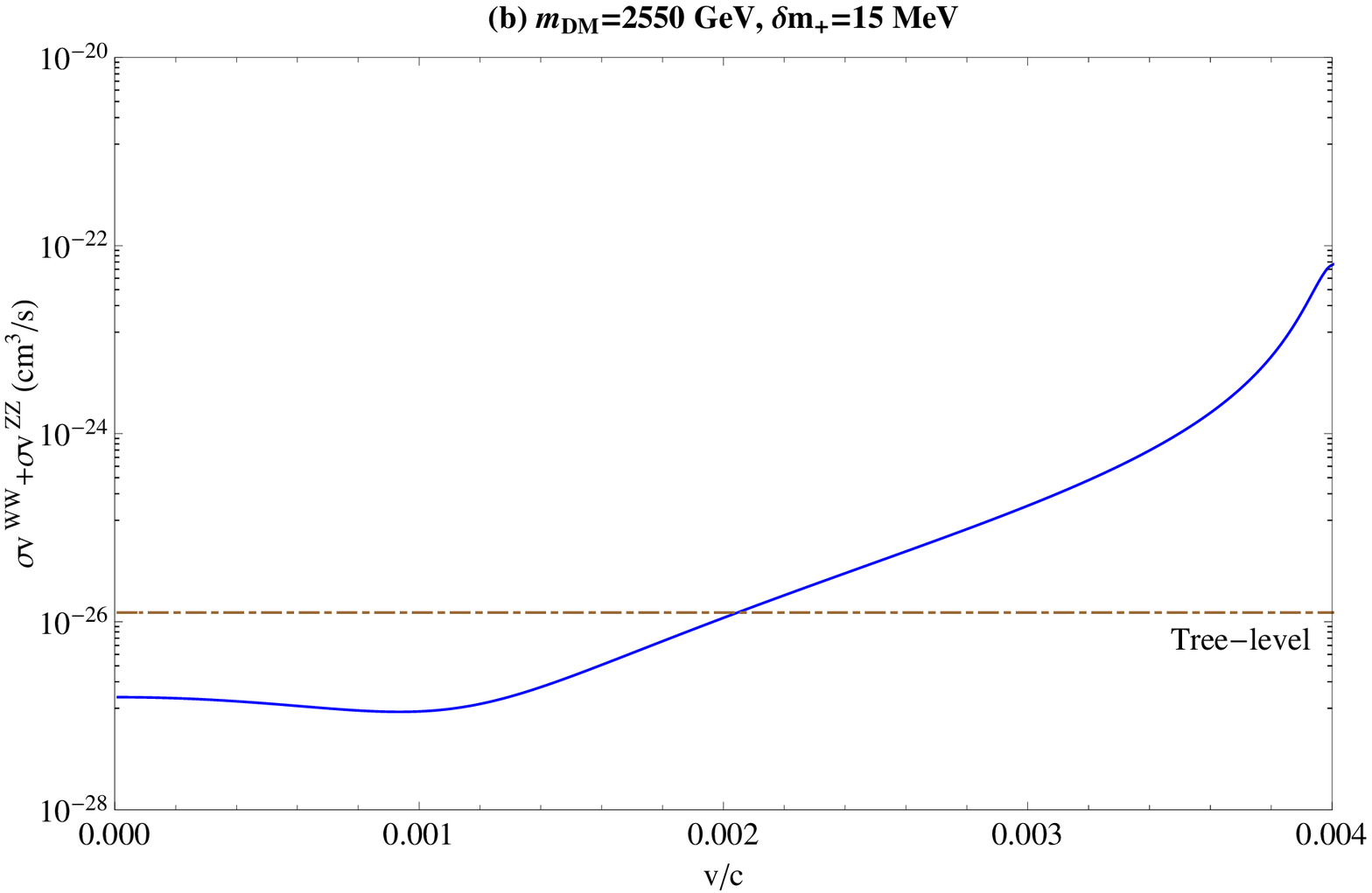}

\includegraphics[width=0.49\linewidth]{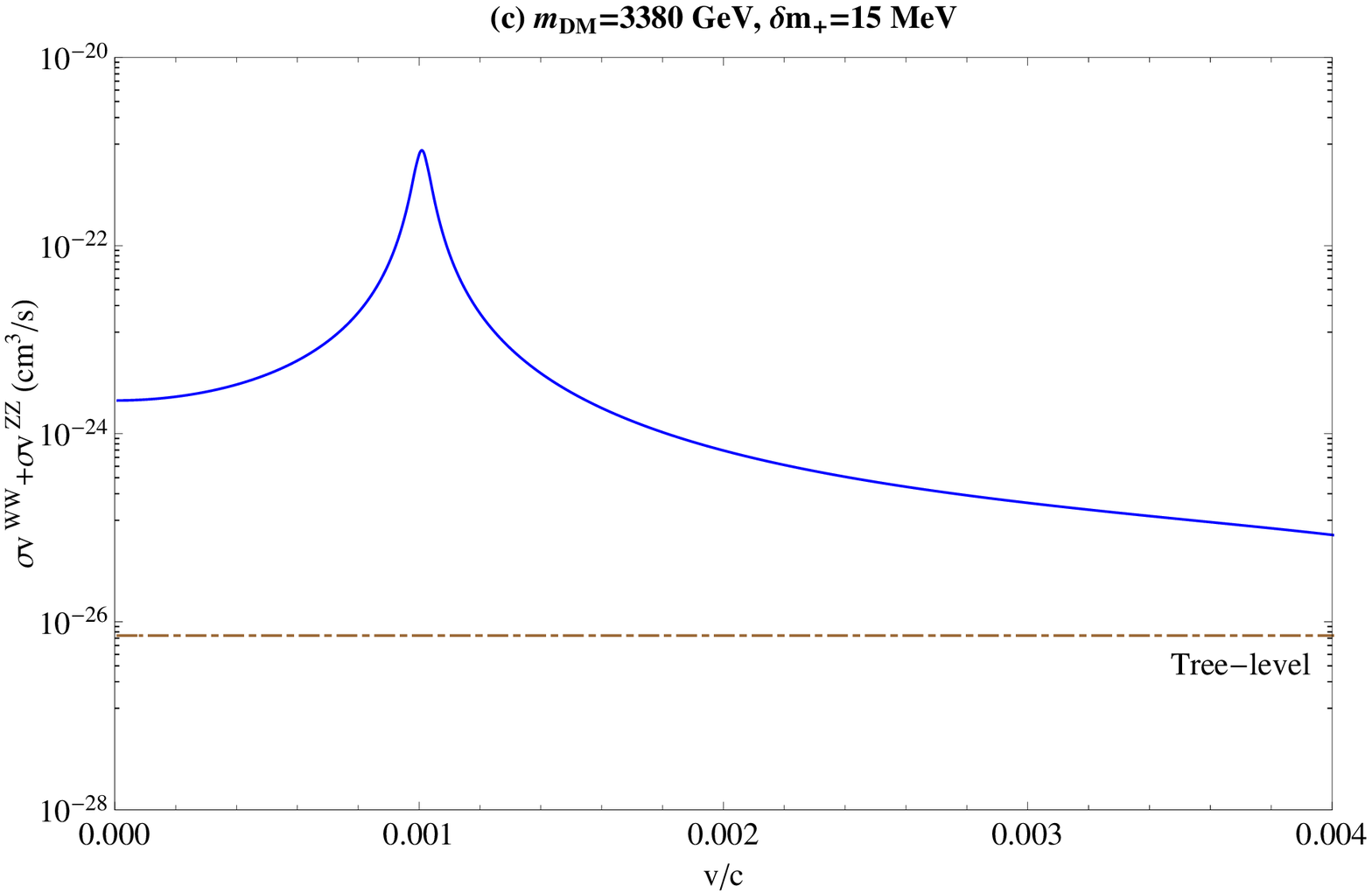}
\end{center}
\caption{Annihilation cross sections  for the wino-like EWDM with
$\delta m_+ =$ 15 MeV as functions of the relative velocity,
$v/c$. For each panel, the used DM mass, $m_{\rm DM}$, corresponds
to the positions of the peaks and the dips of the red dotted line
in Figure \protect\ref{Fig_5-1}. The brown dot-dashed line is the
tree-level cross section for each DM mass.} \label{Fig_5-5}
\end{figure}

As discussed in previous subsections, the SRT resonances occur when
certain conditions are satisfied among the EWDM parameters such as the
electroweak interaction strength, the mass gap and the kinetic energy
(which depends on the DM velocity). So far, we fixed the DM
velocity\footnote{This is also done in other analyses, for instance,
  in Refs.~\protect\cite{hisano03,cirelli07}.} to $v/c = 10^{-3}$.
However, the DM particles in the halo of our Galaxy have a velocity
distribution which is expected, for some particular values of the
parameters, to smooth out the pattern of peaks and dips produced by
the SRT effect. Therefore, for a real physical system it will be
crucial to include the integration over the velocity distribution in
the calculation the EWDM annihilation rate. In Figure \ref{Fig_5-5},
we present the values of $\sigma v^{WW+ZZ}$ in terms of the relative
velocity $v/c$ of the two DM particles for three representative values
$m_{\rm DM} = 1423, 2550$, and 3380 GeV, which correspond to the
positions of the peaks and the dips of the red dotted line in Figure
\ref{Fig_5-1}. As expected, the DM annihilation cross section shows a
dependence on the velocity of the incoming DM particles. For the
analysis of the PAMELA antiproton flux limit in Section
\ref{PAMELA-limit}, we will use the annihilation cross section
obtained after velocity integration according to the following
formula:
\begin{equation}
  \langle \sigma v \rangle=N(v_{esc})
  \int_0^{2 v_{esc}}[\sigma v(v)]v^2exp\left[-\frac{3}{4}
  \left (\frac{v}{v_{rms}} \right)^2\right] \;dv,
\end{equation}
where $v_{rms}$=270 km/s, $v_{esc}$=550 km/s and $N(v_{esc})$ is
the normalization constant.

\subsection{Dependence on $\delta m_N$}

\begin{figure}
\begin{center}
\includegraphics[width=0.65\linewidth]{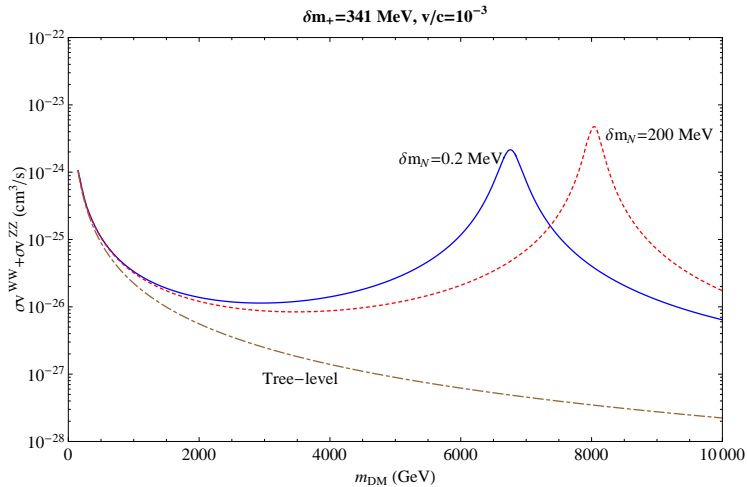}
\end{center}
\caption{Annihilation cross sections for the Higgsino-like EWDM
when $\delta m_N \equiv m_{\chi_1^0}-m_{\chi_0^0} =$0.2 MeV (blue
solid) and 200 MeV (red dotted). The mass difference between the
charged and DM states, $\delta m_+ \equiv
m_{\chi^+}-m_{\chi_0^0}$, is fixed as 341 MeV. The brown
dot-dashed line shows the tree-level cross section.}
\label{Fig_5-6}
\end{figure}

Before closing this section, we will check the dependence of the SRT
effect on the mass splitting between the neutral states. The wino-like
EWDM multiplet has only one Majorana neutral component, but, in
general, EWDM multiplets can have more than one Majorana neutral
component, whose masses are generically different from each other. In
order to see the effect of the neutral mass splitting on the
non-perturbative annihilation rate, we show $\sigma v^{WW+ZZ}$ for the
Higgsino-like EWDM taking two values of the mass splitting: $\delta
m_N \equiv m_{\chi_1^0}-m_{\chi_0^0} =$0.2 and 200 MeV in Figure
\ref{Fig_5-6}. Here the mass difference between the DM and charged
states, $\delta m_+ \equiv m_{\chi^+}-m_{\chi_0^0}$, is fixed to be
341 MeV which is the typical mass splitting arising from the EW
one-loop correction of Higgsino-like EWDM \cite{minimal_dm}. As one
can see in Figure \ref{Fig_5-6}, the position of the peak is shifted
from $m_{\rm DM} \approx 8000$ GeV to $m_{\rm DM} \approx 6800$ GeV
when $\delta m_N$ is changed from 200 MeV to 0.2 MeV. Note that in the
previous Figure \ref{Fig_5-1}, the peak position moved from $m_{\rm
  DM} \approx 2200$ GeV to $m_{\rm DM} \approx 1200$ GeV when $\delta
m_+$ was changed from 166 MeV to 15 MeV. Thus, one can conclude that
the SRT effect for EWDM is much less sensitive to the neutral mass
splitting compared to the splitting between the DM and charged
states. In the following analysis, we will fix $\delta m_N$ to 0.2 MeV
(when applicable) corresponding to a situation for which direct
detection is not excluded by the current experimental data and might
yield a positive signal, as will be shown in Section \ref{sec:direct}.

\section{Direct detection}
\label{sec:direct}

As already pointed out in Section~\ref{General-EWDM}, EWDMs with
non-vanishing hypercharge $Y$ have large couplings to nucleons driven
by the exchange of a $Z$ boson, so their elastic cross sections on
nuclei is excluded by Dark Matter direct--detection experiments
similarly to what typically happens for the simplest realizations of
scenarios where the Dark Matter particle is a massive Dirac neutrino
or a sneutrino. If, however, the elastic cross section for the EWDM is
suppressed, inelastic scattering where the DM particle makes a
transition to a slightly heavier neutral mass eigenstate is possible
\cite{inelastic}. Notice that the wino--like EWDM has a vanishing
hypercharge so is not subject to constraints from direct detection. In
the case of the EWDM with $Y\ne$0, the neutral (Dirac) component of
the multiplet is split into two Majorana particles (the lightest of
which is the DM particle), and thus the elastic cross section vanishes
and only an inelastic transition of the DM particle to the heavier
mass neutral state EWDM$^{\prime}$ is allowed. In particular, among
the cases discussed in the previous sections, this scenario can be
realized for the Higgsino--like EWDM ($T=1/2$, $Y=\pm 1/2$) or for the
triplet EWDM ($T=1$) with $Y=\pm 1$.

\begin{figure}
\begin{center}
\includegraphics[width=0.50\columnwidth]{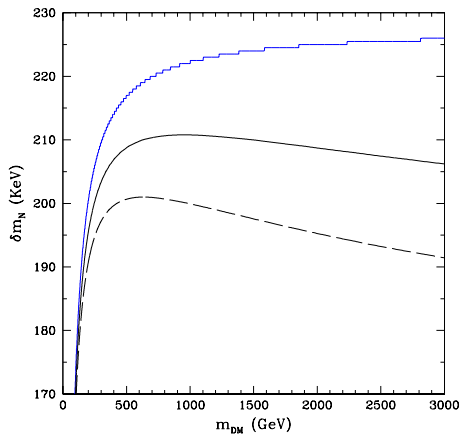}
\end{center}
\caption{Values for the mass splitting $\delta m_{N}$ which are
  detectable and experimentally allowed by the XENON100 direct
  detection experiment \protect\cite{xenon100} as a function of the DM
  mass $m_{DM}$.  The solid lower curve refers to the case $Y=1$ and
  the dashed lower curve to the case $Y=1/2$, while the solid upper
  does not depend on the value of the cross section and is common to
  both cases. When $\delta m_{N}$ is above the upper curve the recoil
  energy is always below the XENON100 lowest-energy threshold
  $E_{th}$=4 keV. On the other hand, values below the lower curves are
  excluded at 90\% C.L. by the 2 nuclear-recoil candidate events
  observed by XENON100 in the range 6.6 keV$< E_r<$ 43.3 keV by
  applying Yellin's maximal gap method \protect\cite{yellin}.}
\label{fig:xenon_inelastic}
\end{figure}

The detection rate of inelastic scattering is suppressed by the
relevant mass splitting $\delta m_{N}$. In particular, for a given recoil
energy $E_{R}$ there exists a minimal velocity for the dark matter
$\beta_{min}$ below which the kinetic energy is not sufficient to
allow the transition to the excited state:
\begin{equation}
\beta_{min}=\sqrt{\frac{1}{2 M_N E_R}}\left (\frac{M_N E_R}{\mu}+\delta m_{N} \right ).
\label{eq:beta_min}
\end{equation}
In the above equation $M_N$ is the nuclear mass and $\mu$ is the
reduced mass of the DM-particle and nucleus. Since the incoming
velocity of DM particles is bounded from above by their escape
velocity $v_{esc}$ in the Galaxy while every DM direct detection
experiment is able to detect DM scatterings only above a given
lower-energy threshold $E_{th}$, detectable values of the
parameter $\delta m_{N}$ are bounded from above, typically a few
hundreds keV depending on the target material and the energy
threshold, while a comparison of the expected signals with the
measured event rates allows to determine a lower bound on $\delta
m_{N}$. Moreover, the EWDM-nucleon inelastic cross section by $Z$
boson exchange is fully determined when the DM mass $m_{DM}$ is
fixed \cite{minimal_dm}:
\begin{equation}
\sigma_{EWDM,nucleon\rightarrow EWDM',nucleon}=
c\frac{G_{\rm F}^2M_{N}^2}{2\pi} Y^2(N - (1-4s_{\rm W}^2) Z)^2\,,
\label{eq:direct_detection_cross_section}
\end{equation}
where the mass splitting between the two neutral states is
neglected, $c=1$ for a fermionic EWDM, and $Z$ and $N$ are the
number of protons and of neutrons in the target nucleus with mass
$M_{N}$. This implies that the allowed range for $\delta m_{N}$
can be plotted as a function of $m_{DM}$.  This is done in Figure
\ref{fig:xenon_inelastic} for the case of the latest constraints
from the XENON100 experiment \cite{xenon100}. In this figure, the
solid lower curve refers to the case $Y=\pm 1$ while the dashed
lower curve corresponds to $Y=\pm 1/2$. As explained above, values
of $\delta m_{N}$ above the upper curve are non--detectable, since
in this case $c\beta_{min} >
v_{max}=v_{esc}+v_{earth}$.\footnote{Notice that $\beta_{min}$ is
defined in the detector rest frame, while $v_{esc}$ is defined in
the Galactic rest frame. For the Galilean boost we have assumed
$v_{earth}$=232 km/s.} This is a purely kinematic constraint that
does not depend on the EWDM--nucleus cross section, so the upper
solid curve in the figure is common to the cases with $Y=\pm 1$
and $Y=\pm 1/2$. On the other hand, the lower curves represent the
90\% C.L. lower bounds on $\delta m_{N}$ obtained by applying
Yellin's maximal gap procedure \cite{yellin} to the spectrum
(consisting of two nuclear recoil candidates at recoil energies
7.1 keV and 7.8 keV) detected by XENON100 in the range 6.6 keV$<
E_r<$ 43.3 \cite{xenon100}. In the calculation, we have assumed
the standard value $\rho_{DM}$=0.3 GeV/cm$^3$ for the DM density
in the neighborhood of the Sun and a Maxwellian velocity
distribution truncated at $v_{esc}$=550 km/s with $v_{rms}$=270
km/s.\footnote{In Ref.~\protect\cite{xenon100}, the DM Region Of
Interest (ROE) 6.6 keV$< E_r<$ 43.3 is used when analyzing the
data with a Profile Likelihood method, but is reduced to 6.6 keV$<
E_r<$ 30.5 keV when applying the maximum-gap method. While this
does not imply a significant change in the limit for elastic
scattering, the inelastic scattering bound is very sensitive to
the upper bound of the ROI. For this reason, we derive our limit
using the whole energy range of the XENON100 measurement.}

As it will be shown in detail, antiproton fluxes are almost
insensitive to the particular choice of $\delta m_N$, while are very
sensitive to the other mass splittings.  Nevertheless, in the
following sections we will fix $\delta m_N$=200 keV (when applicable)
corresponding to a situation for which direct detection is not
excluded by present constraints and might yield a positive signal.

\section{Constraints from antiproton fluxes}
\label{sec:pbar}

Since the EWDM is $SU(2)$ charged, their annihilations are expected to
produce $W/Z$ bosons copiously whenever this channel is kinematically
allowed, leading to a sizeable primary contribution to the antiproton
flux detected by experiments measuring cosmic--rays. The antiproton
primary contribution from DM annihilation must be summed to the
secondary antiproton contribution produced by energetic cosmic rays
impinging on the interstellar medium. Although still affected by
uncertainties, the latter contribution can be calculated in a
relatively reliable way, and is in fair agreement to observation. This
implies that no much room is left for the additional contribution from
DM annihilation, and antiproton data can put constraints on the EWDM,
namely on the EWDM annihilation cross section-times-velocity $\sigma
v$.

\begin{figure}
\begin{center}
\includegraphics[width=0.50\columnwidth]{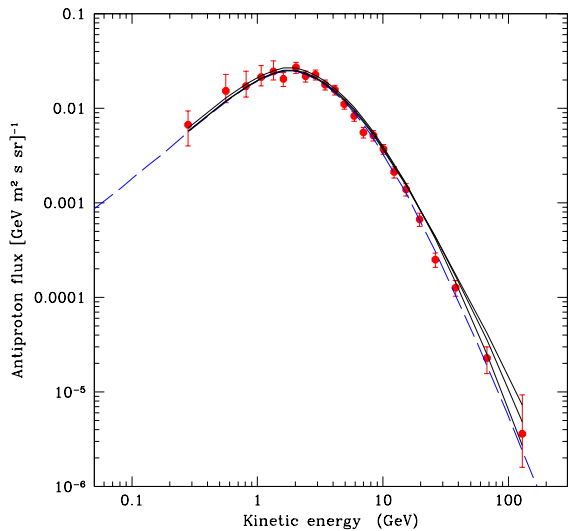}
\end{center}
\caption{The circles show the antiproton top-of-atmosphere flux
  measured by PAMELA \protect\cite{pamela2010} as a function of the
  antiproton kinetic energy. The dashed line represents the secondary
  flux as calculated in \protect\cite{ptuskin} and rescaled by an
  overall factor 0.84. The solid lines show three expected fluxes from
  DM annihilation calculated as described in Section
  \protect\ref{sec:pbar} for $m_{DM}$=200 GeV,500 GeV, and 1 TeV, all
  corresponding to $\chi^2$=44.2.}
\label{fig:antiproton_fit}
\end{figure}

This is shown in Figure \ref{fig:antiproton_fit}, where the circles
represent the top-of-atmosphere antiproton flux as measured by PAMELA
\cite{pamela2010}, while the dashed line is the secondary flux as
calculated in Ref.~\cite{ptuskin}, and rescaled by an overall factor
0.84. In this way, the model fits the data particularly well, with
$\chi^2=12.1$ with 23 degrees of freedom.  For this reason, we adopt
this model as an estimation of the secondary flux.  As far as the
primary antiproton flux is concerned, we have used both the antiproton
yields per annihilation and the propagation model according to
Ref.~\cite{pppc}, adopting for the latter an Einasto density profile
with median values of the propagation parameters.  Notice that, since
the antiproton yields per annihilation corresponding to the two
different final states $WW$ and $ZZ$ are practically
undistinguishable, in the calculation of the expected signal the total
cross section $\sigma^{WW}+\sigma^{ZZ} $ is factorized.

\begin{figure}
\begin{center}
\includegraphics[width=0.50\columnwidth]{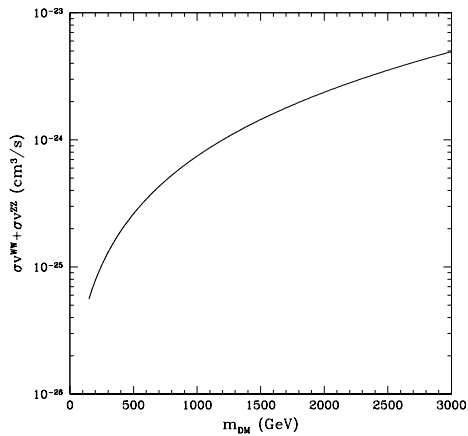}
\end{center}
\caption{Estimation of the 99.5 \% C.L. upper bound on the annihilation cross
  section times velocity $\langle\sigma v \rangle$ as a function of the DM mass
  $m_{DM}$. The values of $\langle\sigma v \rangle$ above the solid line have
  $\chi^2>$44.2 with 23 degrees of freedom when compared to the PAMELA
  data \cite{pamela2010}.}
\label{fig:pamela_sigmav_bound}
\end{figure}

The result of our analysis is shown in
Figures~\ref{fig:antiproton_fit} and \ref{fig:pamela_sigmav_bound}. In
particular, for each choice of $m_{DM}$ and $\sigma v^{WW}+\sigma
v^{ZZ}$ we sum the primary and secondary contributions of the expected
antiproton flux and use the PAMELA data points to calculate a
$\chi^2$, for which we assume an upper bound $\chi^2<$44.2,
corresponding to the 99.5\% C.L. with 23 degrees of freedom. In this
way, we obtain the solid line shown in Figure
\ref{fig:pamela_sigmav_bound}. In Figure \ref{fig:antiproton_fit}, the
three solid lines show the expected antiproton flux for $m_{DM}$=200
GeV, 500 GeV and 1 TeV when $\sigma v^{WW}+\sigma v^{ZZ}$ lies on the
boundary given in Figure \ref{fig:pamela_sigmav_bound}.

\section{PAMELA limit on various EWDM}\label{PAMELA-limit}
\label{sec:pamela_limits}

As shown in Section \ref{sec:pbar}, $W$ and $Z$ bosons produced by the
annihilations of EWDMs provide a sizable contribution to the cosmic
ray antiproton flux measured by PAMELA. Thus, for various EWDM models
we will examine the DM mass ranges satisfying the PAMELA antiproton
flux bound on $\sigma v^{WW+ZZ}$ as obtained in Section
\ref{sec:pbar}. As explained in Section \ref{velocity}, the
annihilation cross section of the EWDM shows a DM velocity
dependence. In this section, we will therefore calculate the
annihilation cross section considering the velocity integration
effect, and also, for comparison, present the result with a fixed
velocity, $v/c=10^{-3}$.

\subsection{Higgsino-like EWDM}

\begin{figure}
\begin{center}
\includegraphics[width=0.49\linewidth]{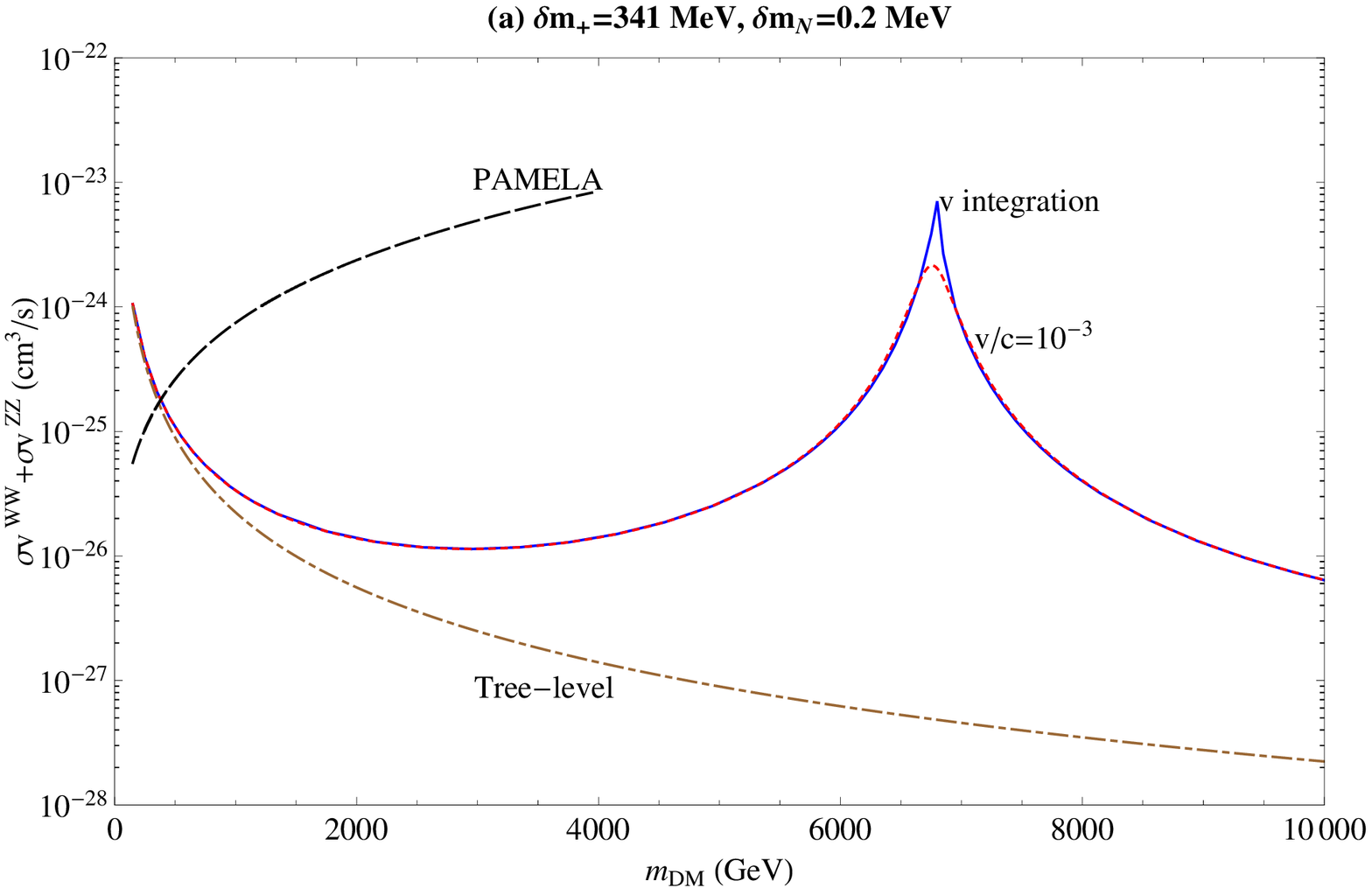}
\includegraphics[width=0.49\linewidth]{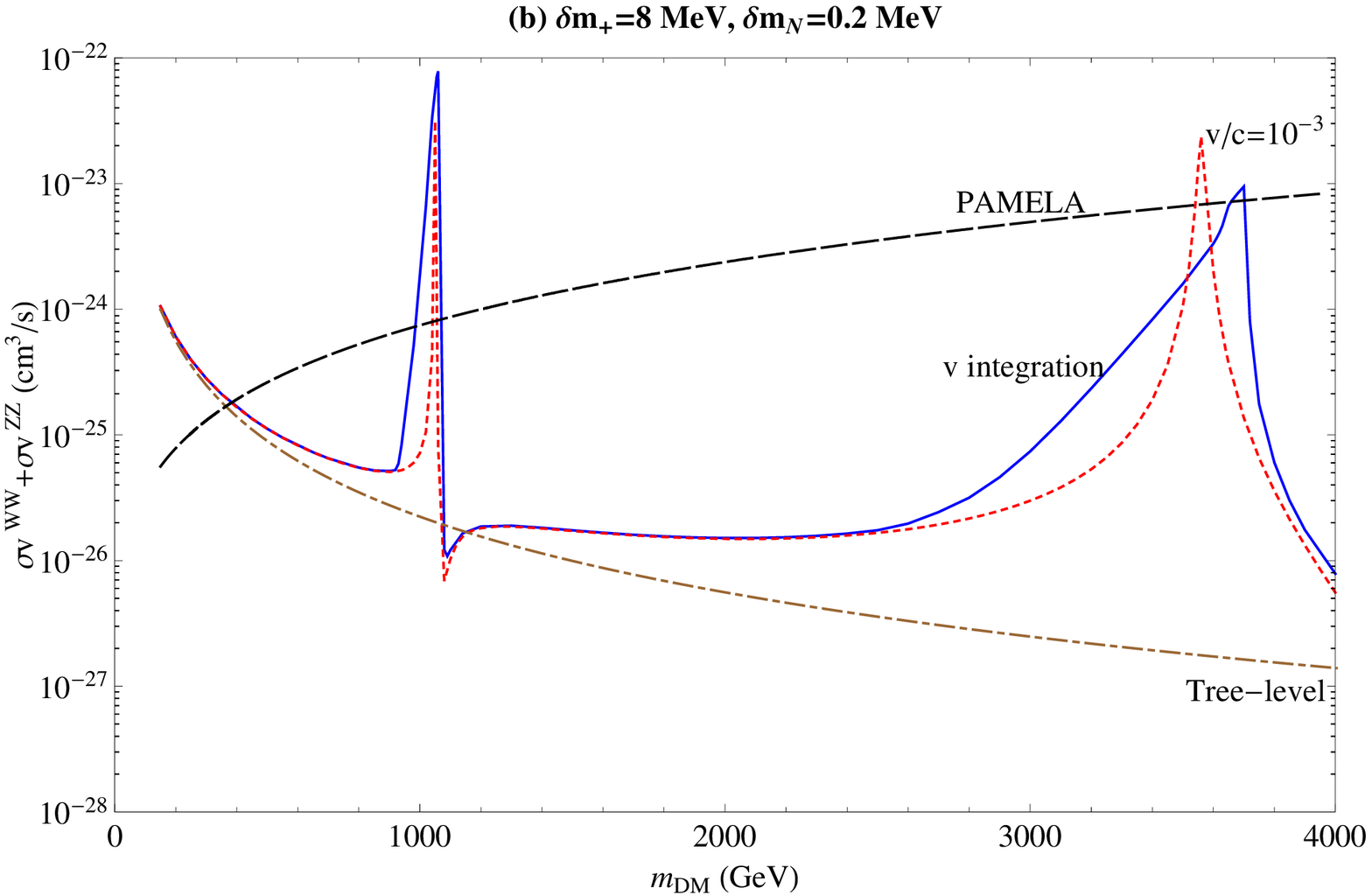}
\end{center}
\caption{Annihilation cross sections to $W^+W^-$ and $ZZ$ for the
Higgsino-like EWDM with $\delta m_+ =$ (a) 341 MeV and (b) 8 MeV,
and $\delta m_N=$ 0.2 MeV. In each panel, the blue solid line is
our final result obtained after velocity integration, while the
red dotted line is the cross section with the fixed velocity,
$v/c=10^{-3}$. The black long-dashed line shows the upper limit
obtained from the PAMELA antiproton flux data analysis in Section
\protect\ref{sec:pbar}. The brown dot-dashed line is the
tree-level cross section.} \label{Fig_6-1}
\end{figure}

We first show the results for the Higgsino-like (doublet with $Y=\pm
1/2$) EWDM which is the smallest multiplet. In Figure \ref{Fig_6-1},
we present the annihilation cross sections (blue solid lines) for two
representative values $\delta m_+ =$ 341 and 8 MeV as a functions of
$m_{\rm DM}$, where the velocity integration effect is included. For
comparison, we also show our results for a fixed relative DM velocity
$v/c=10^{-3}$ (red dotted lines).  As already explained, in this plot
and in the following ones the neutral mass splitting $\delta m_N$ is
taken to be 0.2 MeV, when applicable. For the higher mass splitting
$\delta m_+ = 341$ MeV (a), we extended the DM mass range up to 10 TeV
to see the first Sommerfeld peak. If we further extended the mass
range, Ramsauer-Townsend dips would appear as well.  On the other
hand, for the smaller mass gap $\delta m_+ = 8$ MeV (b), the SRT
resonances appear in the smaller DM mass range as discussed in Section
3.

As shown in the figures, the velocity integration smooths out the
peaks and dips and changes the positions of the peaks in (b). With
only the tree-level cross section, the region $m_{\rm DM} \lesssim
364$ GeV is ruled out by the current PAMELA data. However, the
annihilation cross section is enhanced by the SRT and consequently
the excluded region is a bit extended to $m_{\rm DM} \lesssim 382$
GeV in the case of the typical charged mass splitting for the
Higgsino-like EWDM, $\delta m_+ =$ 341 MeV \cite{minimal_dm}. In
the case of a smaller mass splitting, the PAMELA limit can
constrain also small bands around the Sommerfeld peaks, in
addition to the low mass region, as can be seen in Figure
\ref{Fig_6-1} (b).

\subsection{Wino-like EWDM}

\begin{figure}
\begin{center}
\includegraphics[width=0.49\linewidth]{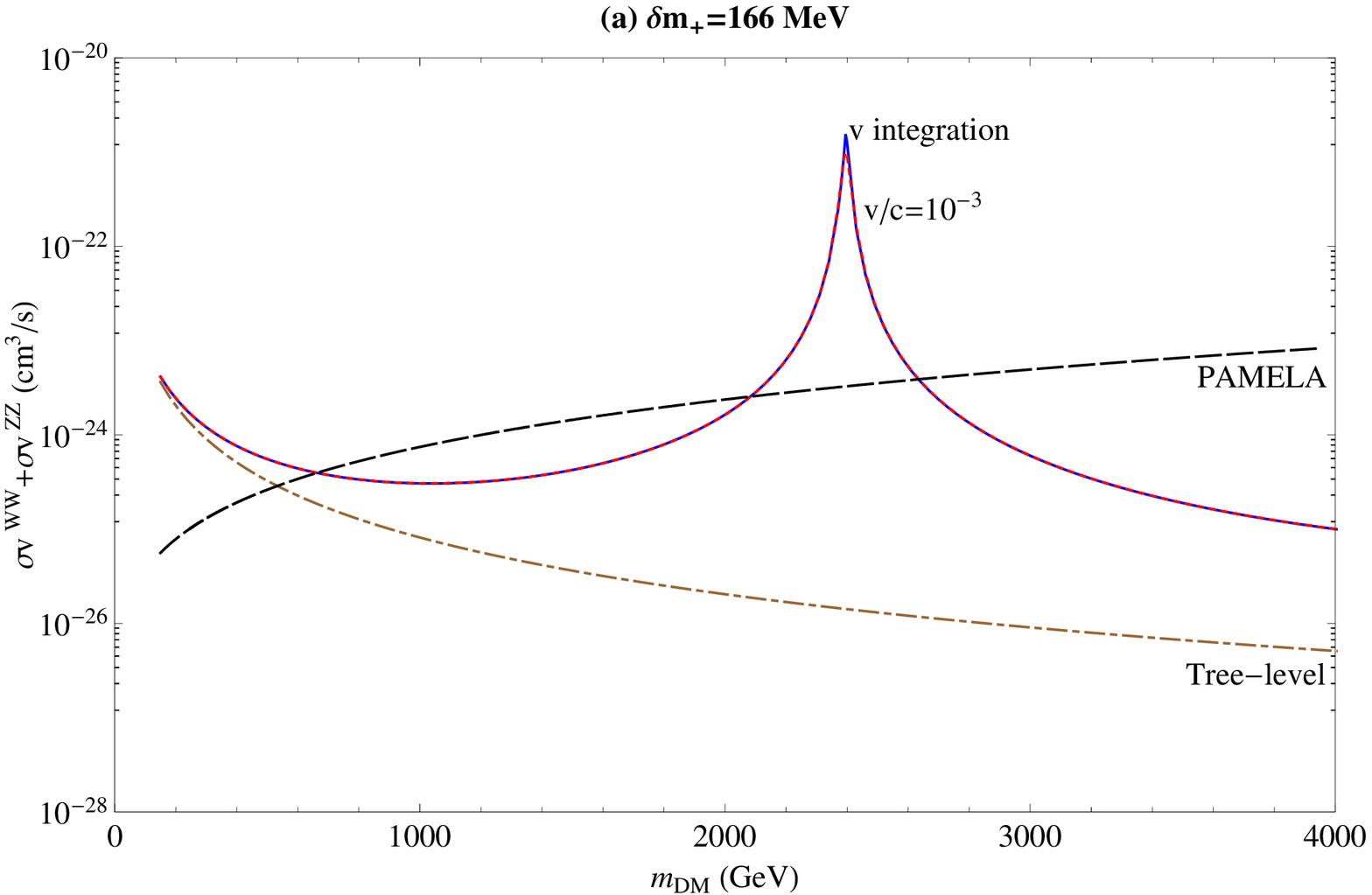}
\includegraphics[width=0.49\linewidth]{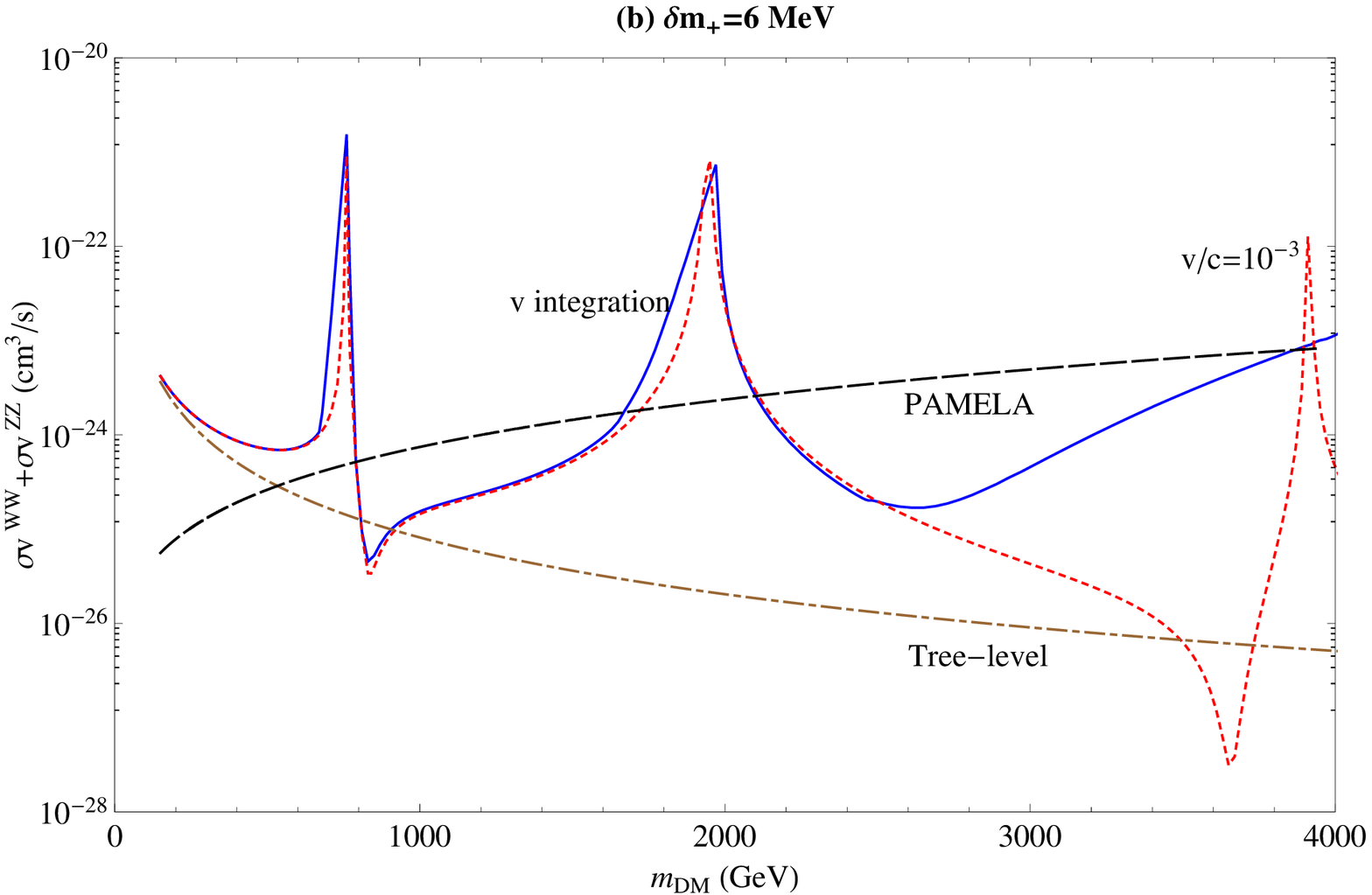}
\end{center}
\caption{Annihilation cross sections to $W^+W^-$ and $ZZ$ for the
wino-like EWDM when $\delta m_+ =$ (a) 166 MeV and (b) 6 MeV. Each
line represents the same thing as in Figure \protect\ref{Fig_6-1}.
} \label{Fig_6-2}
\end{figure}

Figure \ref{Fig_6-2} shows the annihilation cross sections for the
wino-like (triplet with $Y=0$) EWDM in the same way as in the
Higgsino-like EWDM case. The two values $\delta m_+ =$ 166 and 6 MeV
are taken for the analysis, in order to show the dependence on the
charged mass splitting. One can see from Figure \ref{Fig_6-2} that the
velocity integration makes a big change in the case of the smaller
mass gap, erasing out some of the peaks and dips in \ref{Fig_6-2}
(b). While the tree-level result excludes the mass range $m_{\rm DM}
\lesssim 533$ GeV, the non-perturbative effect extends the constrained
region up to $m_{\rm DM} \approx$ 664 GeV for the representative
charged mass splitting of the wino--like EWDM, $\delta m_+ =$ 166
MeV\cite{minimal_dm}. For the smaller mass gap $\delta m_+ = 6$ MeV
the PAMELA limit gets stronger, excluding DM masses below 900 GeV and
also bands around the peaks which are larger compared to the
Higgsino-like EWDM case.

\subsection{Hyper-charged Triplet EWDM}

\begin{figure}
\begin{center}
\includegraphics[width=0.49\linewidth]{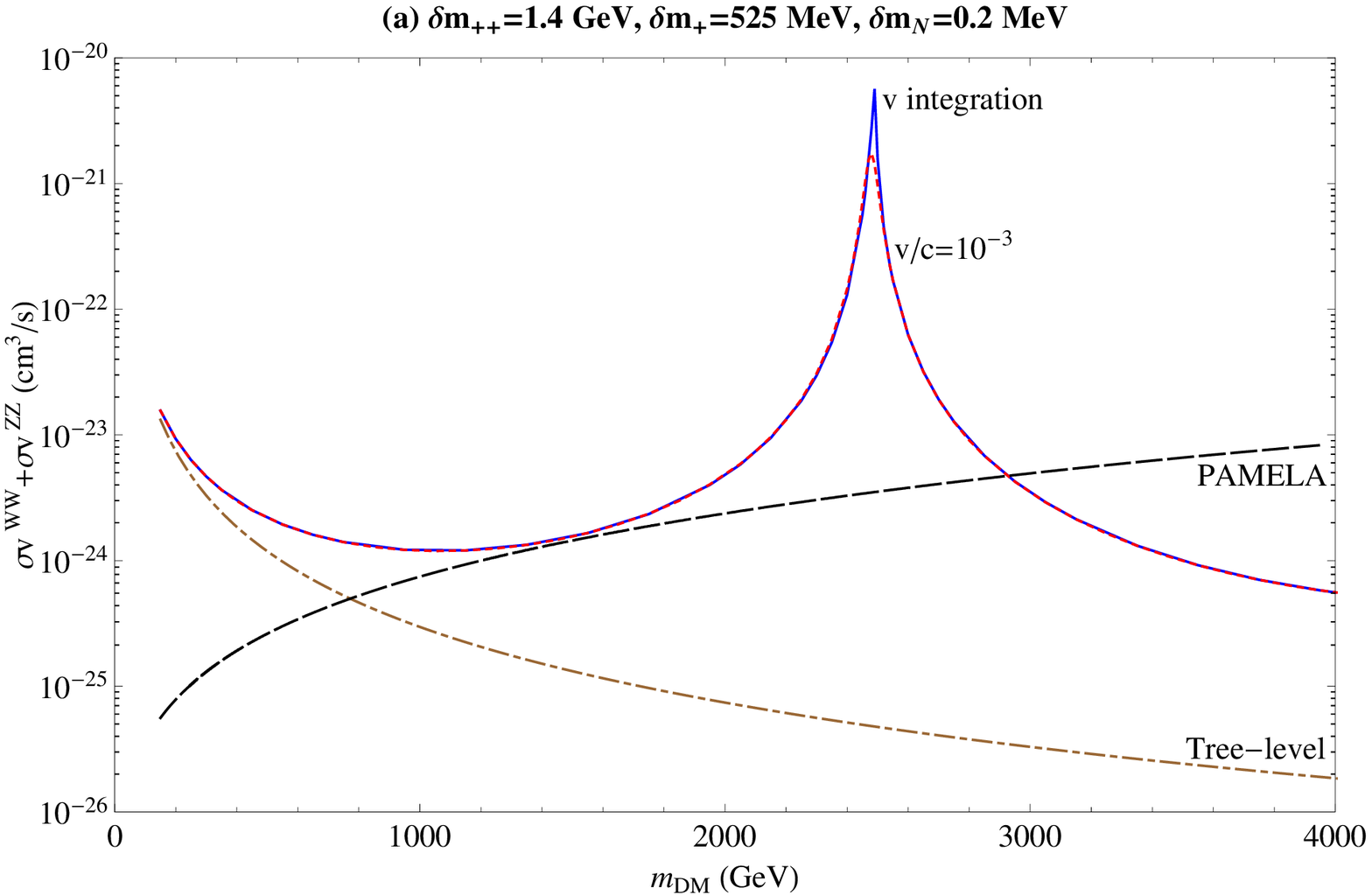}
\includegraphics[width=0.49\linewidth]{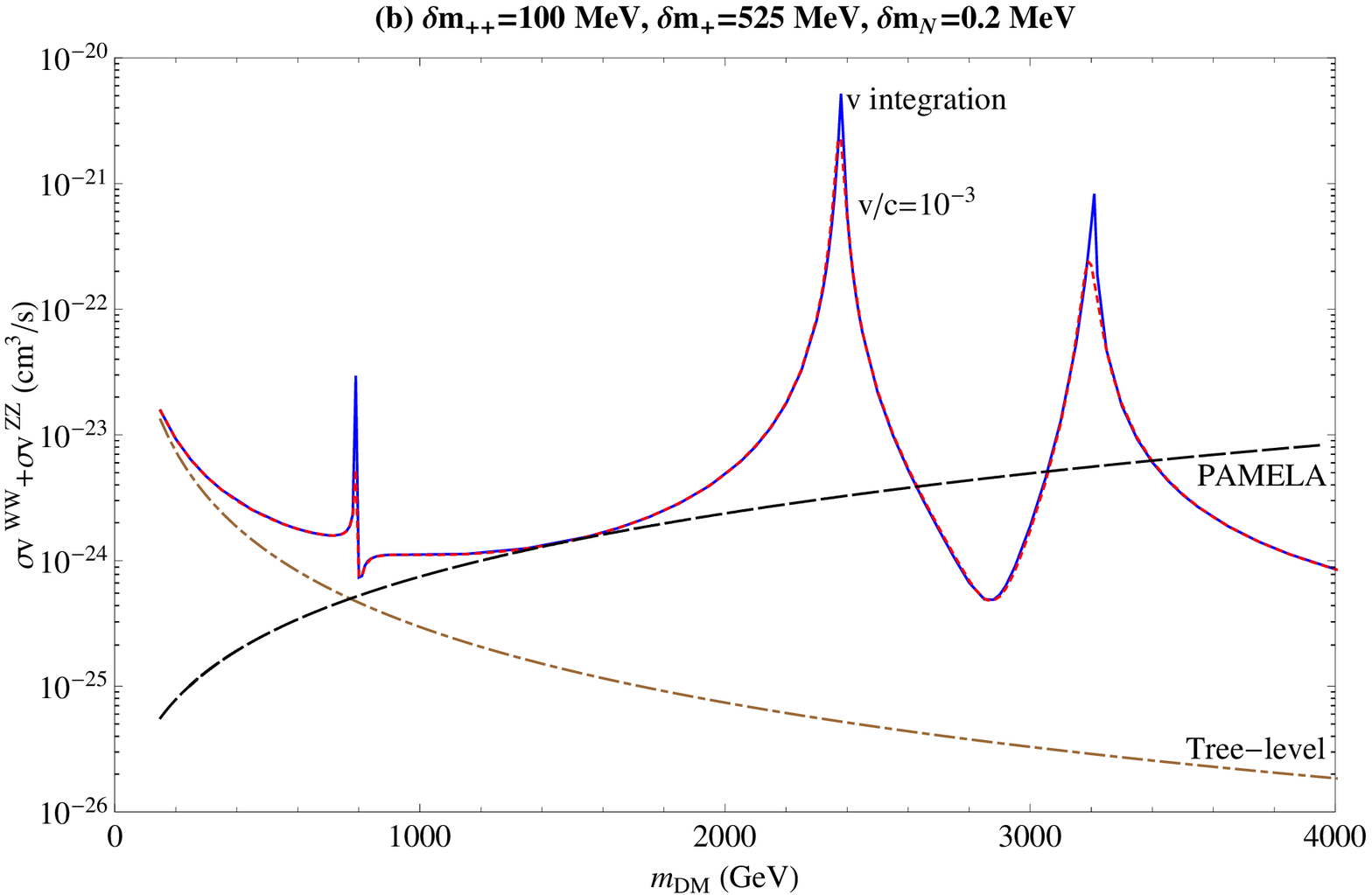}

\includegraphics[width=0.49\linewidth]{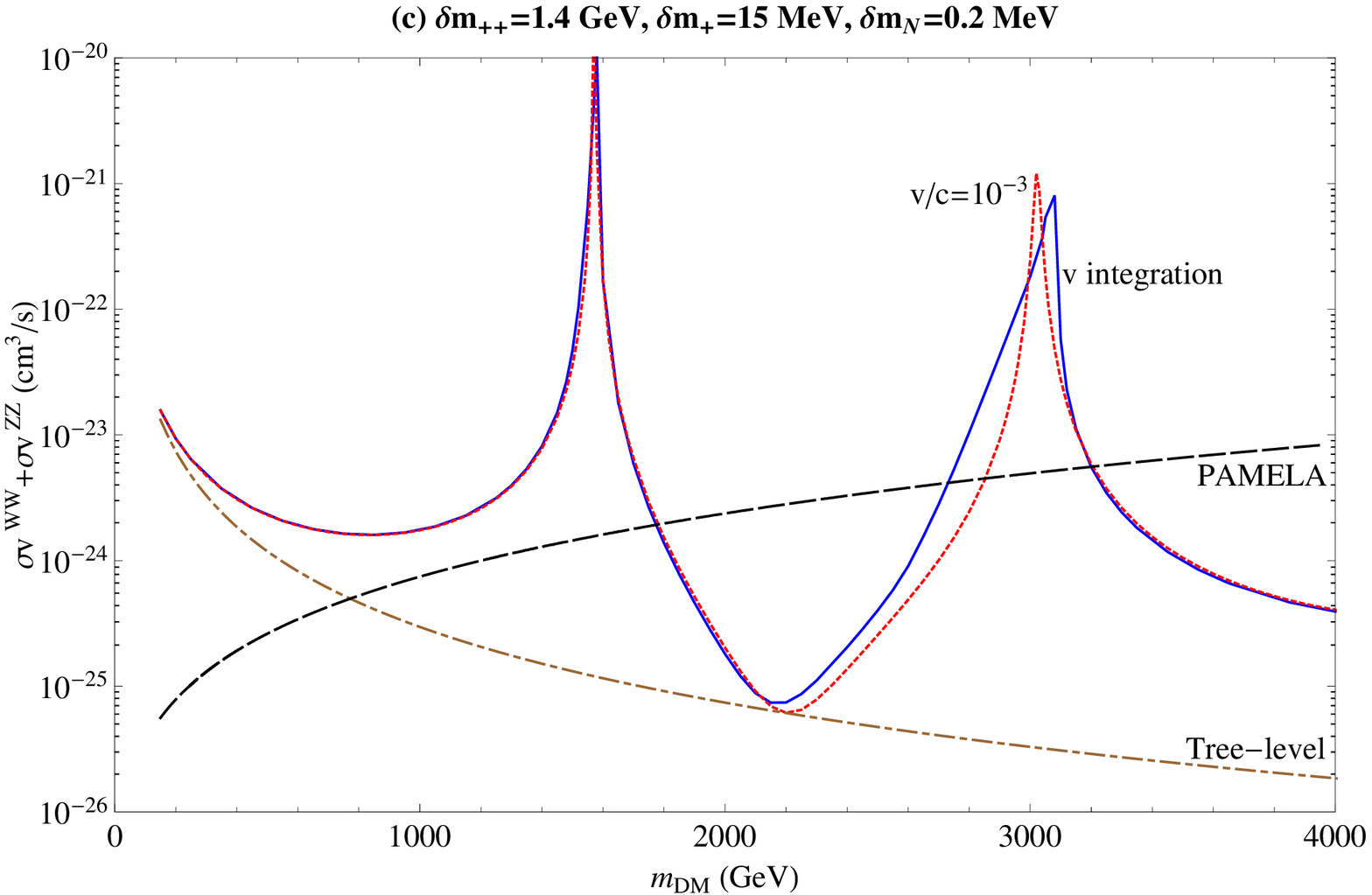}
\includegraphics[width=0.49\linewidth]{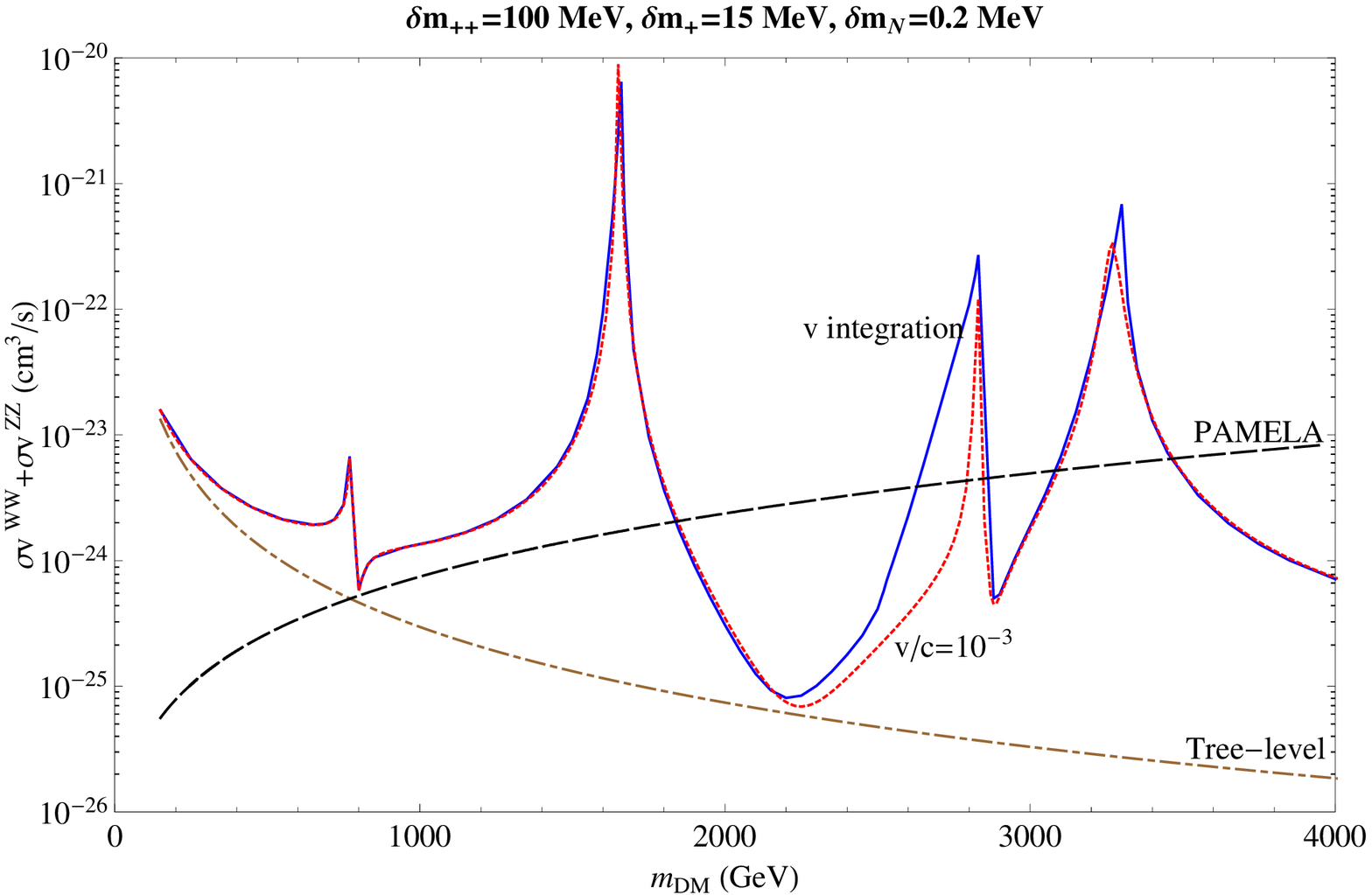}
\end{center}
\caption{Annihilation cross sections to $W^+W^-$ and $ZZ$ for the
triplet EWDM with $Y=\pm1$. The used values of $\delta m_{++, +,
N}$ are shown in each panel. Each line represents the same thing
as in Figure \protect\ref{Fig_6-1}.} \label{Fig_6-3}
\end{figure}

Now let us consider a more complicated case, the triplet EWDM with
$Y=\pm1$ which has one doubly charged, one singly charged, and two
neutral components. Thus, the DM component $\chi_0^0$ can have
three mass splittings: $\delta m_{++} \equiv
m_{\chi^{++}}-m_{\chi_0^0}$, $\delta m_+ \equiv
m_{\chi^+}-m_{\chi_0^0}$, and $\delta m_N \equiv
m_{\chi_1^0}-m_{\chi_0^0}$.  Figure \ref{Fig_6-3} presents the
annihilation cross sections for $(\delta m_{++}, \delta m_+) =
(1400, 525), (100, 525), (1400, 15),$ and $(100, 15)$ MeV fixing
$\delta m_N = 0.2$ MeV. Note that $\delta m_{++} =$ 1400 MeV and
$\delta m_+ =$ 525 MeV correspond to the typical mass splittings
due to the EW one-loop corrections \cite{minimal_dm}.

One can see the dependence of the SRT effect on the doubly charged
mass splitting $\delta m_{++}$ comparing the two cases $(\delta
m_{++}, \delta m_+) = (1400, 525)$ and $(100, 525)$ MeV in Figures
\ref{Fig_6-3} (a) and (b), while the dependence on the singly charged
mass splitting $\delta m_+$ can be understood comparing the two cases
$(\delta m_{++}, \delta m_+) = (1400, 525)$ and $(1400, 15)$ MeV in
Figures \ref{Fig_6-3} (a) and (c).  In particular, one can see that a
larger number of peaks and dips appears when $\delta m_{++}$ is
reduced by a factor 14 compared to the case when $\delta m_{+}$ is
reduced by a factor 35. This shows that the SRT effect is more
sensitive to $\delta m_{++}$ than to $\delta m_{+}$ due to the
stronger EM interaction of multiply-charged states.  In addition,
Figure \ref{Fig_6-3} (d) shows the combined effect of the changes of
$\delta m_{++}$ and $\delta m_+$ shown separately in Figures
\ref{Fig_6-3} (b) and (c).  The hyper-charged triplet EWDM has
stronger EW interactions compared to the Higgsino-like or wino-like
EWDM and thus exhibits a stronger SRT effect. As a consequence, the
excluded mass region reaches about 3 TeV for the typical mass gaps of
$\delta m_{++} =$ 1400 MeV and $\delta m_+ =$ 525 MeV, while the
tree-level limit is as low as about 800 GeV as shown in Figure
\ref{Fig_6-3} (a). It is interesting to see that regions of small DM
mass are allowed for smaller mass gaps, as shown in Figures
\ref{Fig_6-3} (b,c,d) due to the RT effect. One would expect to find
more regions of lower DM mass allowed for lower mass gaps.

\section{Conclusions}
\label{sec:conclusions}

In the present paper, we have discussed the non-perturbative effects
occurring in the annihilation cross section of an ``Electro-Weak Dark
Matter''(EWDM) particle belonging to an $SU(2)_L\times U(1)_Y$
multiplet, when the splittings between the DM state mass and that of
the other charged or neutral component(s) of the multiplet are treated
as free parameters. In particular, we have considered a vector-like
(Dirac) doublet with $Y=\pm 1/2$ (Higgsino-like), a (Majorana) triplet
with $Y=0$ (wino-like) and a vector-like (Dirac) triplet with
$Y=\pm1$. In all these examples, an ad hoc symmetry has to be imposed
for the stability of EWDM, and we have allowed for an unspecified
non-standard cosmology for the generation of the right dark matter
relic density, since the thermal abundance of EWDM is typically below
that required by observation unless its mass is in the multi--TeV
range. Moreover, in the case of EWDM charged under $U(1)_Y$, severe
constraints from direct detection searches exist on the elastic cross
section off nuclei. However, these limits can be circumvented in
presence of a sufficiently large mass splitting (of the order of 0.2
MeV) in the Dirac dark matter fermion, so that only inelastic
scattering is allowed and kinematically suppressed.

As a result of our analysis, it is shown that EWDM exhibits not only
the usual Sommerfeld enhancement of the cross section with resonance
peaks at particular values of the dark matter mass, but also a
suppressed cross section for particular choices of the parameters. The
latter phenomenon is a realization of the ``Ramsauer-Townsend effect''
observed in low-energy electron scattering off gas atoms.
Moreover, we have shown that the EWDM mass for which non-perturbative
effects become important is particularly sensitive to the mass
splittings between the dark matter and the charged components of the
EW multiplet, and is driven below the TeV scale when this mass
splitting is reduced to a few MeV. In particular, we have shown that
when the mass splitting gets smaller the transition of the dark matter
particles to electrically charged states is made easier, and it is the
electromagnetic long-range interaction between these charged states
which is responsible both of the Sommerfeld enhancement of the cross
section and of the Ramsauer-Townsend suppression, even when the dark
matter mass is not much larger than the EW gauge boson masses. Notice
that only mass splittings larger than 100 MeV, induced by EW radiative
corrections, have been considered so far in the literature, so that,
before our analysis, the Sommerfeld effect had been discussed only in
the context of multi-TeV scale dark matter.

Based on the results explained above, we have then used available
experimental constraints on the exotic component of the antiproton
flux in cosmic rays to put constraints to the EWDM parameter
space. Since non-perturbative effects depend on the velocity of the
dark matter particles, we have calculated the annihilation cross
section considering the effect of the convolution of the cross section
with the velocity distribution of the dark matter particles in the
Galaxy, showing that in some cases this can smear out or significantly
modify the pattern of Sommerfeld-Ramsauer-Townsend peaks and dips
obtained for a fixed value of the velocity. Typically, we have found
constraints on the EWDM mass ranging between a few hundred GeV to a
few TeV, depending on the specific EWDM realization and on the values
of the mass splittings. However, we have also found that, by an
appropriate choice of the mass splittings, sometimes the
Ramsauer-Townsend suppression in the annihilation cross section can
allow to recover narrow intervals at lower values of the EWDM mass.

Finally, in the case of EWDM charged under $U(1)_Y$ we have found that
the phenomenology described above is not particularly sensitive to the
mass splitting between the two neutral Majorana states. As a
consequence, this mass difference can be chosen so that the inelastic
cross section of the EWDM off nuclei is allowed by present direct
detection constraints and at the same time is within the reach of
future experiments.

\medskip

\acknowledgments
S.S. acknowledges support by the National Research Foundation of Korea
(NRF) with a grant funded by the Korea government (MEST)
no. 2011-0024836.



\end{document}